# Accuracy of citation data in Web of Science and Scopus

Nees Jan van Eck   Ludo Waltman

*{ecknjpvan, waltmanlr}@cwts.leidenuniv.nl*
Centre for Science and Technology Studies, Leiden University, Leiden (The Netherlands)

**Abstract**
We present a large-scale analysis of the accuracy of citation data in the Web of Science and Scopus databases. The analysis is based on citations given in publications in Elsevier journals. We reveal significant data quality problems for both databases. Missing and incorrect references are important problems in Web of Science. Duplicate publications are a serious problem in Scopus.

**Conference topic**
Data accuracy and disambiguation

**Introduction**
Citation relations between publications are the cornerstone of many bibliometric analyses. For this reason, the availability of accurate citation data is essential for high-quality bibliometric studies. In this paper, we present a large-scale analysis of the accuracy of citation data in Web of Science (WoS) and Scopus, the two most important multidisciplinary bibliometric databases. Our work continues earlier research on this topic (e.g., Buchanan, 2006; Franceschini, Maisano, & Mastrogiacomo, 2013, 2015; García-Pérez, 2010; Olensky, Schmidt, & Van Eck, 2016). The analysis that we present focuses on citations given in publications that appeared in journals published by Elsevier.

The accuracy of citation data can be studied from two perspectives. On the one hand, we can study the extent to which the reference lists of publications are properly represented in the reference data in a bibliometric database. On the other hand, we can study the degree to which a bibliometric database, based on the reference data it contains, manages to correctly identify citation relations between publications indexed in the database. The first problem is about the accuracy of reference data, while the second problem is about the accuracy of citation matching. Our primary focus in this paper is on studying the accuracy of reference data, although we will also provide some insight into the accuracy of citation matching.

**Data**
We used the Elsevier ScienceDirect Article Retrieval API to obtain the reference lists of all Elsevier publications that appeared in the period 1987–2016 (except for a small share of publications not included in the subscription of our university). The reference lists were obtained in XML format. We refer to the data set that we obtained in this way as the Elsevier data set. Publications without references were excluded from this data set.

For each publication in the Elsevier data set, we tried to find corresponding publications in WoS and Scopus. In the case of WoS, we focused on publications of the document types *article* and *review* published in the period 1987–2016 and indexed in the Science Citation Index Expanded, the Social Sciences Citation Index, or the Arts & Humanities Citation Index. In the case of Scopus, we considered publications of the document types *article*, *review*, and *conference paper* published in the period 1996–2015. Publications from 2016 were not taken into account, because most of these publications are still missing in the most recent version of the Scopus database that we have available internally at our center. Publications from before 1996 were not taken into account because of the limited coverage of these publications in Scopus.

To link publications in the Elsevier data set to publications in WoS and Scopus, we first attempted to match publications based on DOI. If no DOI-based match could be obtained, we tried to match publications based on the combination of the name of the first author, the publication year, the volume number, and the first page number. A match was required for all four fields. In the case of matching based on the name of the first author, only the last name and the first initial of the author were taken into account.

Publications in the Elsevier data set that have been linked to publications in WoS and Scopus are referred to as linked publications. The analyses presented in this paper are restricted to these linked publications. Figure 1 shows both for WoS and for Scopus the number of linked publications per year.

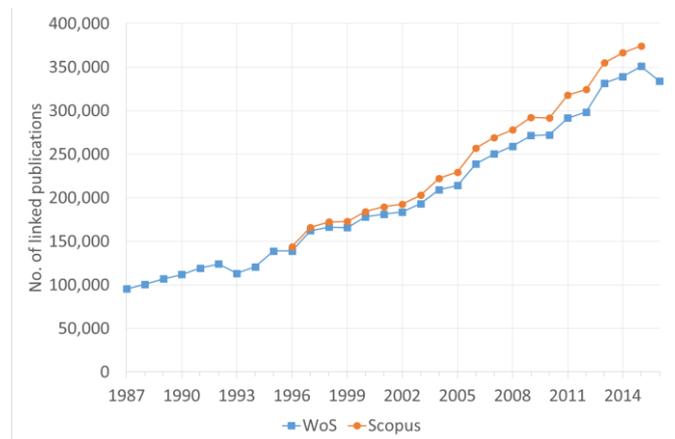

**Figure 1. Time trend in the number of linked publications.**

The next step was to match the references in the linked publications in the Elsevier data set with publications in WoS and Scopus. References in the Elsevier data set do not include a DOI. Matching was therefore done based on the combination of the name of the first author, the publication year, the volume number, and the first page number. References for which a match could be made are referred to as linked references.

Table 1 reports some statistics summarizing the results of linking the Elsevier data set with WoS and Scopus.

**Table 1. Statistics on the linking of the Elsevier data set with WoS and Scopus.**

|  | *WoS* | *Scopus* |
|---|---|---|
| Time period | 1987–2016 | 1996–2015 |
| Document types | article, review | article, review, conference paper |
| No. of linked publications | 6,063,087 | 5,006,165 |
| No. of references in Elsevier data set | 206,540,477 | 172,125,831 |
| No. of references in WoS/Scopus | 203,349,407 | 170,153,108 |
| No. of linked references | 135,559,190 | 84,391,846 |

**Results**

*Analysis based on the number of references in a publication*

We start by making a comparison for the linked publications of the number of references in the Elsevier data set and the number of references in WoS and Scopus. In this comparison, we do not make use of the linking of references discussed above. We consider only the number of references in a publication.

We distinguish between linked publications for which the number of references in WoS or Scopus is equal to, larger than, or smaller than the number of references in the Elsevier data set. A fourth class are linked publications that have no references at all in WoS or Scopus. Table 2 reports the share of the linked publications in WoS and Scopus belonging to the different classes. Time trends are presented in Figure 2.

**Table 2. Classification of linked publications based on their number of references.**

|  | *WoS* | *Scopus* |
|---|---|---|
| Equal no. of references | 77.2% | 96.4% |
| More references | 2.7% | 1.2% |
| Fewer references | 19.3% | 1.2% |
| No references | 0.8% | 1.2% |

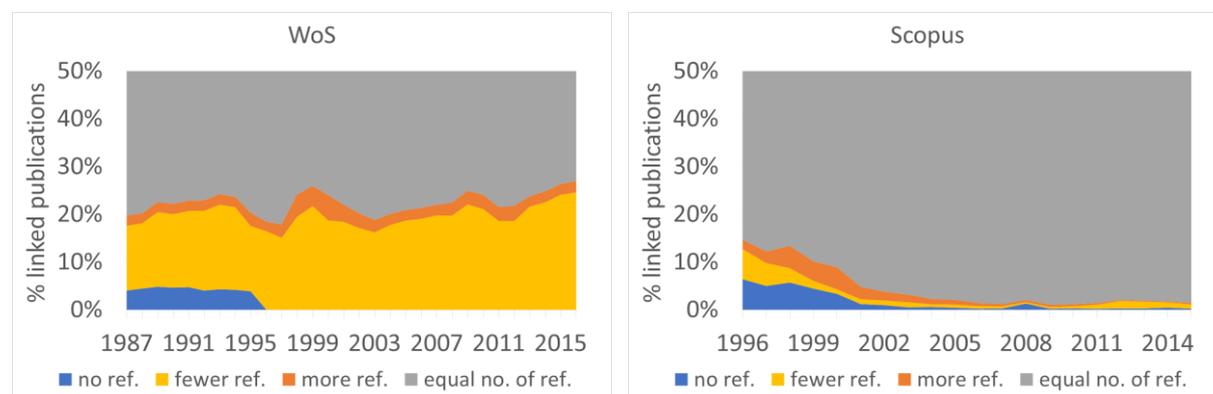

**Figure 2. Time trend in the classification of linked publications based on their number of references.**

Figure 2 shows that in recent years there are no or almost no linked publications without references in WoS and Scopus. In the case of Scopus, for almost all linked publications the number of references in Scopus is the same as in the Elsevier data set. The situation is quite different in the case of WoS. A relatively small share of the linked publications have more references in WoS than in the Elsevier data set. Moreover, a large share of the linked publications have fewer references in WoS than in the Elsevier data set.

To better understand why some linked publications have more references in WoS than in the Elsevier data set, we manually examined ten randomly selected cases. It turns out that a single reference in the Elsevier data set sometimes refers to multiple cited works. It seems that WoS aims to split up such a reference into multiple references, each of them referring to a single cited work. Based on the ten cases that we examined, we have the impression that the approach taken by WoS to split up references does not always give good results. In some cases, however, it works very well. We for instance found a publication (DOI: 10.1016/j.jorganchem.2015.04.038) that includes the following reference:

> J. Gauss, Chem. Phys. Lett. 191 (1992) 614-620. J. Gauss, J. Chem. Phys., 99 (1993) 3629-3643.; A. Schäfer, H. Horn and R. Ahlrichs, J. Chem. Phys., 97 (1992) 2571-2577.

This is seen as a single reference in the Elsevier data set. In WoS, this reference has been correctly split up into the following three references:

> GAUSS J, 1992, CHEM PHYS LETT, V191, P614
> GAUSS J, 1993, J CHEM PHYS, V99, P3629
> SCHAFER A, 1992, J CHEM PHYS, V97, P2571

In Scopus, the reference has not been split up. Instead, the reference has been truncated. In Scopus, it refers only to a single cited work. The other two cited works have simply been left out.

Of course, we also need to better understand why there are so many linked publications that have fewer references in WoS than in the Elsevier data set. The analysis reported below provides more insight into this issue.

*Analysis based on linked references*

We now present a second analysis. This analysis is based on the linked references identified using the approach described in the Data section. For each linked reference, we searched for the corresponding citation relation in WoS or Scopus. In the case of WoS, we used the citation relations identified by a citation matching algorithm developed at our center. We did not have access to the 'official' citation relations identified by the database producer, but our own citation relations have been shown to compare favorably with these 'official' citation relations (Olensky et al., 2016). In the case of Scopus, on the other hand, we did use the 'official' citation relations identified by the database producer.

Of the 136 million linked references in the WoS case (see Table 1), 0.9% did not have a corresponding citation relation. In the Scopus case, 1.2% of the 84 million linked references did not have a corresponding citation relation. Figure 3 shows the time trend in the percentage of linked references for which there is no corresponding citation relation in WoS or Scopus.

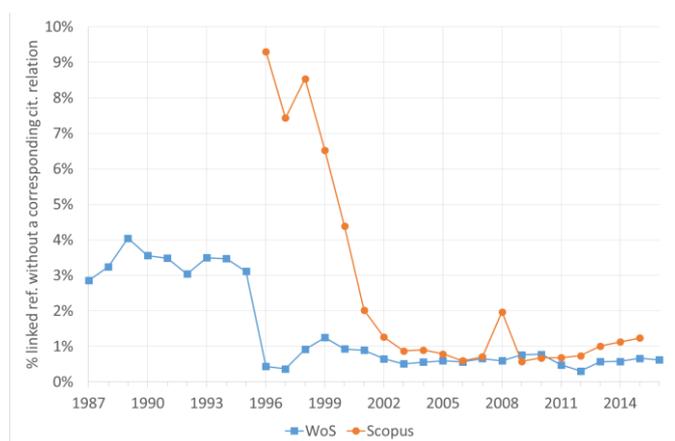

**Figure 3. Time trend in the percentage of linked references for which no corresponding citation relation was found in WoS or Scopus.**

As can be seen in Figure 3, for both databases the percentage of linked references without a corresponding citation relation has decreased significantly over time. In recent years, in the case of Scopus, about 1% of the linked references do not have a corresponding citation relation. This percentage is somewhat lower for WoS. The difference may at least partly be explained by the use of an advanced citation matching algorithm in the case of WoS.

Why do some linked references have no corresponding citation relation in WoS or Scopus? To answer this question, we manually examined a number of randomly selected linked references for which there is no corresponding citation relation. We focused on linked references in publications that appeared in 2015.

In the case of WoS, 0.7% of the linked references in publications from 2015 do not have a corresponding citation relation. Of the 125,583 linked references without a corresponding citation relation, we examined a random sample of 60 cases. It turns out that the 60 cases can be classified into the following four categories:

- Missing reference (33 cases, 55.0%): The reference is missing in WoS (see also Buchanan, 2006; García-Pérez, 2010). Our examination revealed that in some cases multiple references are missing in the same publication. We found one case (DOI: 10.1016/j.virol.2015.02.016) in which the first part of the reference list of a publication is entirely missing. Interestingly, this turned out to coincide exactly with the references listed on one specific page in the PDF file of the publication, suggesting that even in 2015 indexing of Elsevier publications in WoS was still partly based on PDF files.

  The problem of missing references explains why in the analysis based on linked publications discussed above we found that a significant share of the linked publications have fewer references in WoS than in the Elsevier data set.

- Error in reference (16 cases, 26.7%): There is an error in the reference in WoS, such as an incorrect publication year or volume number.

- Incorrect reference (10 cases, 16.7%): The reference has been replaced by a completely different reference (sometimes referred to as a 'phantom reference'; García-Pérez, 2010). The latter reference has some similarity with the former one (e.g., name of the first author, publication year, and perhaps volume number or first page number), but apart from this it is completely different. Some examples of incorrect references are presented in Table 3. We consider incorrect references to be highly problematic, since they result in serious distortions in citation records.

- No problem (1 case, 1.5%): There is no problem. A closer examination showed that for this linked reference there actually does exist a corresponding citation relation, but this citation relation had been overlooked by our algorithms.

**Table 3. Examples of incorrect references in WoS. Elements that the original reference and the incorrect reference in WoS have in common are shown in bold.**

| *Reference in WoS* | *Reference in original publication* |
|---|---|
| **WANG J**, **2006**, CHINESE CHEM LETT, V**17**, P**49** | **J. Wang**, J.K. Carson, M.F. North, D.J. Cleland, Int. J. Heat Mass Transfer 49 (17) (**2006**) 3075–3083. |
| **KANBER B**, **2013**, CEREBROVASC DIS S2, V35, P**21** | **Kanber B**, Hartshorne TC, Horsfield MA, Naylor AR, Robinson TG, Ramnarine KV. Dynamic variations in the ultrasound gray-scale median of carotid artery plaques. Cardiovasc Ultrasound **2013**a;11:**21**. |
| **ZHANG K**, **2014**, IEEE T PATTERN ANAL, V1, P1 | **K. Zhang**, H. Chen, G. Wu, K. Chen, H. Yang, High expression of SPHK1 in sacral chordoma and association with patients' poor prognosis, Med. Oncol. 31 (11) (**2014**) 247. |

In the case of Scopus, 1.2% of the linked references in publications from 2015 do not have a corresponding citation relation. Of the 73,598 linked references without a corresponding citation relation, we examined a random sample of 30 cases. The 30 cases can be classified into the following three categories:

- Missing reference (6 cases, 20.0%): The reference is missing in Scopus. In each of the six cases that we examined, we found that in fact the entire reference list of the citing publication is missing.
- Duplicate publications (9 cases, 30.0%): The reference is available in Scopus, but there is a problem related to duplicate publications (Valderrama-Zurián, Aguilar-Moya, Melero-Fuentes, & Aleixandre-Benavent, 2015). There exist multiple records in Scopus for the publication cited in the reference. When creating citation relations, it is therefore not clear which record should be used for the cited publication. In the nine cases that we examined, we found that the Scopus citation matching algorithm had created a citation relation using an inferior record for the cited publication. The record is incomplete, while another more complete record could have been used as well.
- Citation matching problem (15 cases, 50.0%): The reference is available in Scopus and a citation relation can be created, but for some reason the Scopus citation matching algorithm has failed to do so.

As mentioned above, the version of the Scopus database that we used in our analysis is not entirely up to date. We therefore also examined the 30 cases discussed above in the online version of Scopus. Interestingly, we found that 13 of the 30 cases have been corrected in the online version of Scopus, which seems to suggest that Scopus data quality has improved significantly during the past year.

**Conclusions**

Citation data suffers from inaccuracies both in WoS and in Scopus. However, the inaccuracies are of a quite different nature. Missing references are a quite significant problem in WoS. The problem of incorrect references is even more serious. This problem needs to be fixed urgently. In Scopus, the citation matching algorithm seems to need improvement. Moreover, duplicate publications (Valderrama-Zurián et al., 2015) represent an important data quality problem in Scopus that requires serious attention.

Our analysis has focused on citations given in publications in Elsevier journals. These publications represent a significant share of all publications indexed in WoS and Scopus. However, it is not clear to what extent our findings generalize to publications from other publishers.